\documentclass[aps,pre,reprint,superscriptaddress,amsmath,amssymb,showpacs]{revtex4-1}
\usepackage{graphicx}
\usepackage{bm}
\usepackage{color}

\begin{document}

\title{Nonequilibrium Steady State 
Driven by a Nonlinear Drift Force}%
\author{Chulan Kwon}%
\email{ckwon@mju.ac.kr}
\affiliation{Department of Physics, Myongji University, Yongin, Gyeonggi-Do, 449-728, Republic of Korea}
\author{Ping Ao}
\affiliation{Shanghai Center for Systems Biomedicine and Department of Physics, Shanghai Jiao Tong University, Shanghai, 200240, China}
\date{\today}
\begin{abstract}
We investigate the properties of the nonequilibrium steady state for the stochastic system driven by a nonlinear drift force and influenced by noises which are not identically and independently distributed. The nonequilibrium steady state (NESS) current results from a residual part of the drift force which is not cancelled by the diffusive action of noises. From our previous study for   the linear drift force the NESS current was found to circulate on the equiprobability surface with the maximum at a stable fixed point of the drift force. For the nonlinear drift force, we use the perturbation theory with respect to the cubic and quartic coefficients of the drift force. We find an interesting potential landscape picture where the probability maximum shifts from the fixed point of the drift force and, furthermore, the NESS current has a nontrivial circulation which flows off the equiprobability surface and has various centers not located at the probability maximum. The theoretical result is well confirmed by the computer simulation. 
\end{abstract}
\maketitle

\section{ Introduction }
Statistical mechanics has revealed many interesting properties for nonequilibrium systems. The fluctuation theorem (FT) was first discovered for 
a deterministic nonequilibrium system driven by a shearing force~\cite{evans, evans-searles1,
gallavotti1,gallavotti2}. The FT was also shown to govern a wide class of stochastic nonequilibrium systems~\cite{crooks1,kurchan1,lebowitz1,maes,
hatano-sasa}. The violation of the fluctuation and dissipation relation is also an important characteristics for nonequilibrium systems~\cite{cugliandolo,eyink,bellon,harada-sasa,chetrite1}. The existence of nonzero current is a main consequence observed in the nonequilibrium steady state (NESS) closely related with the violation of the detailed balance (DB) which have been studied up to date~\cite{prost,doering,derrida1,bodineau1,kat,bodineau2,derrida2,knp1}. 

There are several sources to drive a system into nonequilibrium such as an external driving~\cite{garnier,douarche1}, a contact with multiple heat or particle reservoirs~\cite{derrida1,bodineau1,visco,bodineau2,derrida2,komatsu}, a non-conservative force~\cite{kurchan1, kat,turitsyn,chetrite2,knp1}, a time-dependent perturbation for external parameter~\cite{jarzynski1,crooks2,mazonka,zon,taniguchi,angel,knp2}, etc. We recently studied the case in which nonequilibrium is driven by the combination of two sources, the nonconservative force and multiple noises that are not identically and independently distributed~\cite{kat, knp1}. There may be a stochastic process where thermodynamic quantities such as energy, heat, and work are not defined, most likely in biology \cite{zhu}. One of us (P. A.) suggested a way to transform a stochastic differential equation for position like variables to that for position-momentum pairs in zero mass limit~\cite{ao1}. In one dimension it is equivalent to the usual overdamped limit while it is not the case in high dimensions, giving rise to a new stochastic process not derived by the Ito or Stratonovich type calculus~\cite{yin, ao2, ao3}. 

It is yet an open problem how to construct the probability density function (PDF) for the NESS which is the counter part of the Boltzmann factor for the equilibrium case. There have been only a few examples where the PDF is rigorously found~\cite{derrida1,kat,tailleur,turitsyn,chetrite2}. The so called potential landscape function is defined as minus the logarithm of the PDF, which corresponds to  inverse temperature times the potential or energy for equilibrium systems. Much efforts were made in earlier works by Graham {\it et al.}~\cite{graham} to develop a general formalism to construct the potential landscape function, which is not complete. The potential landscape with probability flux drawn gives a useful and intuitive picture particularly in biology, chemistry, and evolutionary science~\cite{wang}. In this paper we obtain the potential landscape function and the circulating NESS current for the system driven by a nonlinear drift force in the presence of noises not identically and independently distributed. We find a very peculiar landscape picture that cannot be expected easily from a given drift force. It is the generalization of our previous study for the linear drift force~\cite{kat}. For the nonlinear drift force, which cannot be treated rigorously, an approximation method is required. We will use the perturbation theory based on the exact result for the linear drift force as unperturbed basis, which is valid in the low noise limit. The recent theory of large deviation, solving the instanton solution to extremize the rate functional, can also be applied for the low noise cases~\cite{tailleur,bertini}. We expect our approach to be efficient provided that the unperturbed result is known. The two approaches will essentially give rise to an equivalent result.    

We organize the paper as follows. In Sec.~\ref{section2} we discuss general aspects on the NESS for the stochastic dynamics driven by the nonlinear drift force. In Sec.~\ref{section3} we develop the first order perturbation theory for the nonlinear drift force and obtain the correlation functions by using the Fourier transformation of the Langevin equation into the frequency space. In Sec.~\ref{section4} we find the potential landscape function in compliance with the correlation functions found. In Sec.~\ref{section5} we give an explicit expression for the NESS current. In Sec.~\ref{section6} we investigate an example for the motion in a double well potential in the presence of correlated noises. In Sec.~\ref{section7} we present the numerical calculation for the perturbation theory and the simulation for the example. In Sec.~\ref{section8} we summarize our main results.

\section{General Aspects\label{section2}}

We consider the Langevin equation:
\begin{equation}
\dot{{\mathbf x}}=\mathbf{f}({\mathbf x})+\bm{\xi}~. \label{stochastic}
\end{equation}
where the state ${\mathbf x}=(x_1,\ldots,x_n)^{\tau}$, the drift force ${\mathbf f}=(f_1,\ldots,f_n)^{\tau}$, and the noise $\bm\xi=(\xi_1,\ldots,\xi_n)^{\tau}$ are $n$-dimensional vectors. The superscript $\tau$ denotes the transpose of vector and matrix. We consider $\mathbf{x}$ to have the even parity in time reversal like position variables. ${\bm\xi}(t)$ is taken to be Gaussian with zero mean and variance
$\langle \bm{\xi}(t)\bm{\xi}^{\tau}(t')\rangle=2\mathsf{D}\delta(t-t')$. $\mathsf{D}$ is a $n\times n$ real symmetric matrix, called the diffusion matrix, and is positive definite. We consider the noise components $\xi_i$'s to be not identically and independently distributed. Then $\mathsf{D}$ is in general not proportional to the unit matrix, which is a possible realization of the contact with multiple heat reservoirs. We consider the localized system in which $\mathbf{f}$ has stable fixed points and goes to $\infty$ as $|\mathbf{x}|\to\infty$.

The PDF $\rho(\mathbf{x},t)$ associated with the Langevin equation (\ref{stochastic}) satisfies the Fokker-Planck equation
\begin{equation}
{\partial\over\partial t}{\rho}(\mathbf{x},t)=\bm{\nabla}\cdot (-\mathbf{f}+\mathsf{D}\cdot\bm{\nabla})
\rho(\mathbf{x},t)~.
\end{equation}
The steady state is reached when $\partial\rho/\partial t=0$. The PDF for the steady state reads
\begin{equation}
\rho(\mathbf{x})\propto \exp\left(-\Phi(\mathbf{x})\right)~.
\end{equation}
We call $\Phi(\mathbf{x})$ the potential landscape function. It is equal to the inverse temperature times the energy for the equilibrium case. 

The probability current reads $\mathbf{j}=(\mathbf{f}+\mathsf{D}\cdot\bm{\nabla}\Phi)\rho$. Equilibrium is defined as steady state with the DB. The DB for even parity variables is found to hold if $\mathbf{j}=0$~\cite{risken}, i.e. $\mathbf{f}=-\mathsf{D}\cdot\bm{\nabla}\Phi$. In this situation, the drift force is exactly cancelled by the diffusive force $\mathsf{D}\cdot\bm{\nabla}\Phi$ given by noises. The NESS is characterized by nonzero $\mathbf{j}$ which is divergenceless since $\dot{\rho}=0$. For the localized system, that we are considering in this paper,  circulation is the only way for divergenceless current. On the other hand, for the extended system it is directed through the system from one to another boundary. There were interesting works on the noise induced directed current in an one-dimensional extended system with the periodic boundary condition, which explains well the transportation of drugs or molecules in biological systems \cite{prost,doering}. There are common ingredients for the two cases, circulating and directed currents, which we will discuss later.  

Let us define the force matrix $\mathsf{F}$ and the potential matrix $\mathsf{U}$ respectively as
\begin{equation}
F_{ij}=\nabla_{j}f_{i}~,~~U_{ij}=\nabla_{i}\nabla_{j}\Phi~.
\end{equation}
Then the DB condition gives $\mathsf{U}=-\mathsf{D}^{-1}\mathsf{F}$, which leads to the condition:
\begin{equation}
\mathsf{FD}-\mathsf{DF}^{\tau}=0~,\label{detailed-balance}
\end{equation}
which can be obtained by using $\mathsf{U}=\mathsf{U}^{\tau}$.   

The thermodynamic equilibrium is given by $\mathbf{f}=-\gamma\bm{\nabla}E$ for energy $E$ and $\mathsf{D}=(\gamma k_B T)\mathsf{I}$ for unit matrix $\mathsf{I}$. Since $\mathsf{F}=\mathsf{F}^{\tau}$, the DB holds. In fact the DB always holds for one dimensional systems. For higher dimensions there are two factors that make the DB violated. One is that $\mathsf{F}$ is asymmetric, $\mathsf{F}\neq \mathsf{F}^{\tau}$, which is the case in which $\mathbf{f}$ is nonconservative, i.e., not derivable from a scalar function, $\mathbf{f}\not\propto -\bm{\nabla} V$. The other is that $\mathsf{D}\not\propto \mathsf{I}$. Then Eq.~(\ref{detailed-balance}) may not be satisfied. 

With the broken DB, the system will reach the NESS after a long time. Let us define the residual force as
\begin{equation}
\mathbf{f}_{res}=\mathbf{f}+\mathsf{D}\cdot\bm{\nabla} \Phi~,\label{f_res}
\end{equation}
which vanishes for the equilibrium case. The NESS current is given as $\mathbf{j}=\mathbf{f}_{res}\rho$. The condition $\bm{\nabla}\cdot\mathbf{ j}=0$ gives 
\begin{equation}
\bm{\nabla}\cdot \mathbf{f}_{res}-\bm{\nabla} \Phi\cdot \mathbf{f}_{res}=0~. \label{steady-state}
\end{equation}

We briefly summarize our previous study on the case with linear drift force, $\mathbf{f}=\mathsf{F}\cdot\mathbf{x}$, which is known as a high dimensional Ornstein-Uhlenbeck process~\cite{kat}. The PDF is Gaussian and the potential landscape function is given by $\Phi=\mathbf{x}^{\tau}\cdot\mathsf{U}\cdot\mathbf{x}/2$. We found 
\begin{equation}
\mathbf{f}_{res}=-\mathsf{Q}\cdot\bm{\nabla} \Phi~, \label{two-part}
\end{equation}
where $\mathsf{Q}$ is an antisymmetric matrix and independent of $\mathbf{x}$. It satisfies the steady state condition~(\ref{steady-state}). Then the resultant current circulates on the equiprobability surface, which can be shown by noting that $\mathsf{Q}\cdot\bm{\nabla} \Phi$ is perpendicular to $\bm{\nabla}\Phi$. $\bm{\nabla}\Phi=\mathsf{U}\cdot\mathbf{x}$ yields $\mathsf{U}=-(\mathsf{D}+\mathsf{Q})^{-1}\mathsf{F}$. The condition $\mathsf{U}=\mathsf{U}^{\tau}$ gives the matrix equation for the antisymmetric matrix,
\begin{equation}
\mathsf{FQ}+\mathsf{QF}^{\tau}=\mathsf{FD}-\mathsf{DF}^{\tau}~. \label{Q-equation}
\end{equation}
The exact solution was found by using the Jordan transformation for an arbitrary asymmetric matrix $\mathsf{F}$. The antisymmetric matrix $\mathsf{Q}$ is the single measure for the NESS. If $\mathsf{Q\neq 0}$, the DB is violated, as seen in Eqs.~(\ref{detailed-balance}), (\ref{Q-equation}). Recently Filliger and Reimann \cite{filliger} showed there exists a non-vanishing torque perpendicular to a two dimensional heat
engine, which is nothing but a manifestation of the circulating current in our study.

For the nonlinear drift force, however, the antisymmetric matrix is not a sufficient measure for the NESS.  We can write
\begin{equation} 
\mathbf{f}_{res}=-\mathsf{Q}\cdot\bm{\nabla} \Phi+\mathbf{f}_{off}~.
\label{new-decomp} 
\end{equation}
A new term $\mathbf{f}_{off}$ gives a current flowing off the equiprobability surface. Then the drift force is decomposed into three parts:
\begin{equation}
\mathbf{f}=-\mathsf{D}\cdot\bm{\nabla}\Phi-\mathsf{Q}\cdot\bm{\nabla}\Phi+\mathbf{f}_{off}
\end{equation} 
A similar idea of decomposition was used in earlier works by Graham and T\'{e}l \cite{graham}, where they concentrated only on the dissipative part, the first term, in our terminology. A vector potential $\mathbf{A}$ was introduced by Qian \cite{qian} to describe a circular flux
and the global transport in a two dimensional motor protein movement, where $\bm{\nabla}\times\mathbf{A}$ corresponds to the circulating current in our study.

One can see that the probability maximum where $\bm{\nabla}\Phi=0$ shifts from the fixed point where $\mathbf{f}=0$ if $\mathbf{f}^{off}\neq 0$.  It is contrary to a usual observation that the fixed point coincides with the probability maximum, as in the DB case where $\mathbf{f}=-\mathsf{D}^{-1}\cdot\bm{\nabla}\Phi$. $\mathsf{Q}$ may become dependent on $\mathbf{x}$. Then the resultant current  $-(\mathsf{Q}(\mathbf{x})\cdot\bm{\nabla}\Phi)\rho$, flowing on the equiprobability surface, may not be divergenceless. However, the total NESS current with $\mathbf{f}_{off}\rho$ added goes divergenceless and circulates off the equiprobability surface. Interestingly there may appear various centers of circulation, which will be seen in the following sections. 

\section{Perturbation Theory\label{section3}}

Our aim is to find the PDF for the NESS. The nonlinear drift force cannot be treated rigorously for a general $\mathsf{D}$. We use the perturbation theory for the nonlinear force expanded about a fixed point. Let $\mathbf{x}=0$ be the fixed point. Then one can write
\begin{equation}
f_{i}= F^{(0)}_{ij}x_j+G_{ijk}x_{j}x_{k}+H_{ijkl}x_{j}x_{k}x_{l}+\cdots~.
\end{equation}
The Einstein convention is used for repeated indices denoting the summation. The potential landscape function can also be written as
\begin{eqnarray}
\Phi &=& h_i x_i+{1\over 2}\left(U^{(0)}_{ij}+U^{(1)}_{ij}\right)
x_i x_j\nonumber\\
&& +{1\over 3!}\Gamma_{ijk}x_{i}x_{j}x_{k}+{1\over 4!} \Delta_{ijkl}x_{i}x_{j}x_{k}x_{l}+\cdots ~. \label{potential}
\end{eqnarray}

The linear term in the drift force yields the unperturbed basis for the perturbation theory where the unperturbed potential matrix $\mathsf{U}^{(0)}$ is known for $\mathsf{F}^{(0)}$ from our previous study~\cite{kat}. The superscript $(1)$ denotes the perturbation due to the nonlinear drift coefficients. The perturbation theory is expected to be the same as the low noise limit, i.e., the small $\mathsf{D}$ limit, since the PDF is dominant near fixed point in this limit. We then proceed the perturbation theory up to the first order in $G_{ijk}$, $H_{ijkl}$, both treated as of the same order. It is the same sense as in the perturbation theory for the anharmonic effect of the lattice vibration where the cubic and quartic terms in the potential are treated as of the same order in order to give a bound potential.

Note that the linear term $h_{i} x_{i}$ in $\Phi$ is responsible for the shift of the PDF maximum from $\mathbf{x}=0$. The moments of the PDF comprise a set of correlation functions written in terms of the coefficients of $\Phi$. On the other hand we will find the moments directly from the Langevin equation in the Fourier space. Then we can determine the coefficients of $\Phi$ in a selfconsistent manner by comparing the moments found from the two means.

Let $C_{ij\cdots}$ be the connected correlation function $\langle
x_{i}x_{j}\cdots\rangle_{c}$, known as a cumulant. $\langle \cdots\rangle$ denotes the average evaluated by the PDF associated with $\Phi$. For example, $C_{ij}=\langle x_ix_j\rangle-\langle x_i\rangle\langle x_j\rangle$. Up to the first order we
find
\begin{eqnarray}
C_{i} &=& -R_{il}h_{l}- {1\over 2}\Gamma_{lmn}R_{il}R_{mn}~,
\label{first} \\
C_{ij}&=& R_{ij}-[\mathsf{RU}^{(1)}\mathsf{R}]_{ij}
-{1\over 2}\Delta_{klmn}R_{ik}R_{jl}R_{mn}~,
\label{second} \\
C_{ijk} &=& -\Gamma_{lmn}R_{il}R_{jm}R_{kn}~,
\label{third} \\
C_{ijkl}&=& -\Delta_{i'j'k'l'} R_{ii'}R_{jj'}R_{kk'}R_{ll'}~,\label{fourth}
\end{eqnarray}
where we use
\begin{equation}
\mathsf{R}=(\mathsf{U}^{(0)})^{-1}~.
\end{equation}
One can use the standard Feynman diagram technic for vertex tensors $\mathsf{\Gamma}$, $\mathsf{\Delta}$ to obtain the above results.

Using the Fourier transformation one can see that the Langevin equation, the stochastic differential equation in time, turns into the algebraic one for $\mathbf{x}(\omega)$ in frequency space. Up to the first order in $G_{ijk}$, $H_{ijkl}$ we find
\begin{eqnarray}
\lefteqn{x_i(\omega) =\alpha_{ij}(\omega)\Big\{\xi_{j}(\omega)}\nonumber\\
 && +G_{jlm}\int
{d\omega_{1}\over\sqrt{2\pi}}\alpha_{ll'}(\omega_1)\alpha_{mm'}(\omega-\omega_1)
\xi_{l'}(\omega_1)\xi_{m'}(\omega-\omega_1) \nonumber\\
&& +H_{jlmn}\int{d\omega_{1}d\omega_{2}\over
2\pi}\alpha_{ll'}(\omega_{1})\alpha_{mm'}(\omega_{2})\alpha_{nn'}(\omega-\omega_{1}-\omega_{2}) \nonumber\\
&&~~\times\xi_{l'}(\omega_{1})\xi_{m'}(\omega_{2}) \xi_{n'}(\omega-\omega_{1}-\omega_{2}) \Big\}~,
\end{eqnarray}
where we introduce a matrix
\begin{equation}
\mathsf{\alpha}(\omega)= \left[-i\omega \mathsf{I}-\mathsf{F}^{(0)}\right]^{-1}~.
\end{equation}
From this we can find the correlation functions $\langle x_{i}(\omega)x_{j}(\omega') \cdots\rangle$ in frequency space, using
\begin{equation}
\langle \xi_{i}(\omega)\xi_{j}(\omega')\rangle=2D_{ij}\delta(\omega+\omega')~.
\end{equation}
In this calculation $\langle\cdots\rangle$ denotes the average over noises. The connected correlation functions at equal time can then be found by
\begin{equation}
C_{ij\cdots}(t)=\int{d\omega d\omega'\over\ 2\pi} \cdots\langle x_{i}(\omega)x_{j}(\omega')\cdots\rangle_{c} e^{-i(\omega +\omega'+\cdots)t}~.
\end{equation}
The connected correlation functions in frequency space can be found to have a delta function factor, $\delta(\omega+\omega'+\cdots)$. Therefore the connected
correlation functions at equal time are independent of time and coincide with the corresponding correlation functions in the steady state, given in
Eqs.~(\ref{first})--(\ref{fourth}).

After a careful calculation, we find the correlation functions:
\begin{equation}
C_{i}=-2F^{-1}_{ij}G_{jlm}\beta_{lm}(\omega)~, \label{first_time}
\end{equation}
\begin{eqnarray}
C_{ij} &=& \int {d\omega\over \pi}\beta_{ij}(\omega)+12H_{klmn} \int{d\omega_{1}\over
2\pi}\beta_{mn}(\omega_{1})\nonumber\\
&& ~~ \times\int {d\omega\over 2\pi}\Big\{ \alpha_{ik}(-\omega)\beta_{jl}(\omega)+(i\leftrightarrow j) \Big\}~, \label{second_time}
\end{eqnarray}
\begin{eqnarray}
C_{ijk}&=& 8 G_{lmn}\int {d\omega d\omega'\over (2\pi)^2} \Big\{\beta_{im}(\omega)\beta_{jn}(\omega')
\alpha_{kl}(-\omega-\omega')  \nonumber\\
&& +(j\leftrightarrow k)+(i\leftrightarrow k)\Big\}~,\label{third_time}
\end{eqnarray}
\begin{eqnarray}
C_{ijkl}&=& 48 H_{l'abc}\int {d\omega d\omega' d\omega'' \over (2\pi)^3} \Big\{\nonumber\\
&&\beta_{ia}(\omega)\beta_{jb}(\omega')\beta_{kc}(\omega'')\alpha_{ll'}(-\omega-\omega'-\omega'')
\nonumber\\
&&+(k\leftrightarrow l)+(j\leftrightarrow l)+(i\leftrightarrow l)\Big\}~. \label{fourth_time}
\end{eqnarray}
where $(i\leftrightarrow j)$ denotes the term obtained by exchanging indices $i$ and $j$.  We introduce another matrix 
\begin{equation}
\mathsf{\beta}(\omega)= \mathsf{\alpha}(\omega)\mathsf{D}\mathsf{\alpha}^{\tau}(-\omega)~. \label{alpha-beta}
\end{equation}

We can determine the coefficients in $\Phi$ by comparing Eqs.~(\ref{first_time})--(\ref{fourth_time}) with Eqs.~(\ref{first})--(\ref{fourth}). 

\section{Probability Density Function\label{section4}}

Comparing Eqs.~(\ref{second}) and (\ref{second_time}), we can find the potential matrix in zeroth order
\begin{equation}
\mathsf{R}=[\mathsf{U}^{(0)}]^{-1}=\int {d\omega\over\pi} \mathsf{\beta}(\omega)~. \label{U-inverse-integral}
\end{equation}
This gives the antisymmetric matrix in zeroth order,
\begin{equation}
\mathsf{Q}^{(0)}=-\mathsf{F}^{(0)}\int {d\omega\over\pi} (i\omega \mathsf{I}+\mathsf{F}^{(0)})^{-1}D(i\omega I-F^{(0)\tau})^{-1}-\mathsf{D}~, \label{Q-integral}
\end{equation}
where the relation $\mathsf{U}^{(0)}=-(\mathsf{D}+\mathsf{Q}^{(0)})^{-1}\mathsf{F}^{(0)}$ is used. It is very plausible to get the integral representation for the antisymmetric matrix which was found as a complicated series form in our previous work for the linear drift case~\cite{kat}.

For simplicity of notation, let us define
\begin{eqnarray}
\hat\alpha(\omega)&=& \mathsf{U}^{(0)}\alpha(\omega)~,\\
\hat\beta(\omega) &=& \mathsf{U}^{(0)}\beta(\omega)~.
\end{eqnarray}
Comparing Eqs.~(\ref{third}) and (\ref{third_time}), we find the coefficients for the cubic term in $\Phi$:
\begin{equation}
\Gamma_{ijk}=\Xi_{ijk}^{abc}G_{abc}~,
\end{equation}
where
\begin{eqnarray}
\Xi_{ijk}^{abc}&=&-8\int{d\omega d\omega'\over(2\pi)^2}\Big\{{\hat\beta}_{ib}(\omega){\hat\beta}_{jc}(\omega'){\hat\alpha}_{ka}(-\omega-\omega')
\nonumber \\
&& +(j\leftrightarrow k)+(i\leftrightarrow k)\Big\}~. \label{Xi}
\end{eqnarray}

Comparing Eqs.~(\ref{fourth}) and (\ref{fourth_time}), we find the coefficients for the quartic term in $\Phi$:
\begin{equation}
\Delta_{ijkl}=\Psi_{ijkl}^{abcd}H_{abcd}~,
\end{equation}
where
\begin{eqnarray}
\lefteqn{\Psi_{ijkl}^{abcd}=-48\int{d\omega d\omega' d\omega'' \over (2\pi)^3} \Big\{}\nonumber\\
&&{\hat\beta}_{ib}(\omega){\hat\beta}_{jc}(\omega'){\hat\beta}_{kd}(\omega'')
{\hat\alpha}_{la}(-\omega-\omega'-\omega'')\nonumber\\
&&~~~+(k\leftrightarrow l)+(j\leftrightarrow l)+ (i\leftrightarrow l)~\Big\}~. \label{Psi}
\end{eqnarray}

Comparing Eqs.~(\ref{second}) and (\ref{second_time}) in the first order, we find the first order correction for the potential matrix:
\begin{equation}
U^{(1)}_{ij}=-{1\over 2}\left(\Delta_{ijmn}+\Theta_{ij}^{kl}H_{klmn}\right)R_{mn}~, \label{U(1)}
\end{equation}
where
\begin{equation}
\Theta_{ij}^{kl}=12\int{d\omega\over 2\pi}\left\{{\hat\beta}_{il}(\omega){\hat\alpha}_{jk}(-\omega)+(i \leftrightarrow j)\right\}~.\label{Theta}
\end{equation}

Finally, comparing Eq.~(\ref{first}) and (\ref{first_time}), we find the coefficients for the linear term in $\Phi$:
\begin{equation}
h_i=-\left( [(\mathsf{D}+\mathsf{Q}^{(0)})^{-1}]_{il}G_{lmn}+{1\over 2}\Gamma_{imn}\right)R_{mn}~. \label{hi}
\end{equation}
Then we can estimate the shift $\mathbf{x}_{max}$ of the probability maximum from $\mathbf{x}=0$, which is given from $\bm{\nabla} \Phi=0$,
\begin{eqnarray}
[\mathbf{x}_{max}]_i&=& -R_{ij}h_{j}\nonumber\\
&=&\left(-[(\mathsf{F}^{(0)})^{-1}]_{il}G_{lmn}+{1\over 2}R_{il}\Gamma_{lmn}\right)R_{mn}~. \label{shift}
\end{eqnarray}
From Eqs.~(\ref{first}) and (\ref{shift}), we can find  $\langle \mathbf{x}\rangle$ as
\begin{equation}
\langle x_i\rangle=-[(\mathsf{F}^{(0)})^{-1}]_{il}G_{lmn}R_{mn}~. \label{first-mom}
\end{equation}
It is rather surprising that the shift of the probability maximum might not be in the direction of $\langle \mathbf{x}\rangle$.

\section{Nonequilibrium Steady State Current\label{section5}}

We divide the residual force in Eq.~(\ref{new-decomp}) into the zeroth and first order as $\mathbf{f}_{res}=\mathbf{f}_{res}^{(0)}+\mathbf{f}_{res}^{(1)}$. Each term is given as  
\begin{eqnarray}
\mathbf{f}_{res}^{(0)}&=&-\mathsf{Q}^{(0)}\cdot\bm{\nabla}\Phi~,\nonumber\\
\mathbf{f}_{res}^{(1)}&=&\mathbf{f}_{on}^{(1)}+\mathbf{f}_{off}^{(1)}=-\mathsf{Q}^{(1)}\cdot\bm{\nabla}\Phi^{(0)}+\mathbf{f}_{off}~.
\end{eqnarray}
$\mathbf{f}_{on}^{(1)}$ is the term with the first order antisymmetric matrix $\mathsf{Q}^{(1)}$. $\mathbf{f}_{off}$ only appears in the first order, so equals to $\mathbf{f}_{off}^{(1)}$. 

Similarly the NESS current is decomposed as 
\begin{eqnarray}
\mathbf{j}&=&\mathbf{j}^{(0)}+\mathbf{j}^{(1)}=\mathbf{j}^{(0)}+\mathbf{j}_{on}^{(1)}+\mathbf{j}_{off}^{(1)}\nonumber\\
&=&\mathbf{f}_{res}^{(0)}\rho+\mathbf{f}_{on}^{(1)}\rho+\mathbf{f}_{off}^{(1)}\rho
\end{eqnarray}
$\mathbf{j}^{(0)}$ flows on the equiprobability surface. It is divergenceless because $\mathsf{Q}^{(0)}$ is constant in $\mathbf{x}$. $\mathbf{j}^{(1)}_{on}$  also flows on the equiprobability surface, while it may not be divergenceless since $\mathsf{Q}^{(1)}$ generally depends on $\mathbf{x}$. $\mathbf{j}^{(1)}_{off}$ flows off the equiprobability surface. The total current $\mathbf{j}^{(1)}$ in the first order should be divergenceless. The existence of nonzero current is the signal for the NESS.  The existence of the current flowing off the equiprobability surface is resultant from the combination of the nonlinearity of the drift force and the nonuniformity of noises, which is not seen in the linear case.

The velocity field $\mathbf{f}^{(1)}_{res}$ for the total current $\mathbf{j}^{(1)}$ in the first order can be found from 
\begin{equation}
\bm{\nabla}\Phi^{(1)}=-(\mathsf{D}+\mathsf{Q}^{(0)})^{-1}\left(\mathbf{f}^{(1)}-\mathbf{f}^{(1)}_{res}\right)~,\label{gauge-1}
\end{equation}
where $\Phi^{(1)}$ is the first order correction to $\Phi$ and $\mathbf{f}^{(1)}$ is the nonlinear part of $\mathbf{f}$. We find
\begin{eqnarray}
\lefteqn{[\mathbf{f}^{(1)}_{res}]_{i}=[\mathsf{D}+\mathsf{Q}^{(0)}]_{ij}h_{j}}\nonumber\\
&&+[(\mathsf{D}+\mathsf{Q}^{(0)})\mathsf{U}^{(1)}]_{ik}x_{k} \nonumber\\
&&+\left(G_{ikl}+{1\over 2}[\mathsf{D}+\mathsf{Q}^{(0)}]_{ij}\Gamma_{jkl}\right)x_{k}x_{l} \nonumber\\
&&+\left(H_{iklm}+{1\over 3!}[\mathsf{D}+\mathsf{Q}^{(0)}]_{ij}\Delta_{jklm}\right)x_{k}x_{l}x_{m}~. \label{finv}
\end{eqnarray}
The resultant current $\mathbf{j}^{(1)}$ is expected to be divergenceless. 

Note that $\mathbf{f}^{(1)}_{res}$ in the above equation is independent of $\mathsf{Q}^{(1)}$. In fact there is no unique way to determine $\mathsf{Q}^{(1)}$ and $\mathbf{f}_{off}^{(1)}$ simultaneously. There is a gauge invariance under the change: $\mathsf{Q}^{(1)}\to\mathsf{Q}^{(1)}+\delta\mathsf{Q}^{(1)}$, $\mathbf{f}^{(1)}_{off}\to \mathbf{f}^{(1)}_{off}-\mathsf{Q}^{(1)}\cdot\bm{\nabla}\Phi^{(1)}$.
One way is to extract the antisymmetric matrix maximally from $\mathbf{f}^{(1)}_{res}$. 
We expand the perturbed antisymmetric matrix $Q^{(1)}$ as
\begin{equation}
Q^{(1)}_{ij}= q_{ij}+ q_{ijk}x_{k}+q_{ijkl}x_{k}x_{l}+\cdots~,
\end{equation}
where $q_{ij\ldots}=-q_{ji\ldots}$. The part to be extracted from $\mathbf{f}^{(1)}_{res}$ in Eq.~(\ref{finv}) reads
\begin{eqnarray}
-\mathsf{Q}^{(1)}\cdot\bm{\nabla} \Phi^{(0)}&=&-q_{ij}U_{jk}^{(0)}x_k-q_{ijk}U_{jl}^{(0)}x_{k}x_{l}\nonumber\\
&&-q_{ijkl}U_{jm}^{(0)} x_{k}x_{l}x_{m}+\cdots~.
\end{eqnarray}
We find
\begin{equation}
q_{ij}= -{1\over 2}\left(\left[\mathsf{(D+Q^{(0)})U^{(1)}R}\right]_{ij}-(i\leftrightarrow j)\right)~, \label{maximal-Q0}
\end{equation}
\begin{eqnarray}
\lefteqn{q_{ijk}= -{1\over 2}\left(G_{ikl}R_{lj}-(i\leftrightarrow j)\right)}\nonumber\\
&&-{1\over 4}\left([\mathsf{D+Q^{(0)}}]_{im}\Gamma_{mkl}R_{lj}-(i\leftrightarrow j)\right)~, \label{maximal-Q1}
\end{eqnarray}
\begin{eqnarray}
\lefteqn{q_{ijkl}= -{1\over 2}\left(H_{iklm}R_{mj}-(i\leftrightarrow j)\right) }  \nonumber\\
&& - {1\over 12}\left([\mathsf{D+Q^{(0)}}]_{in}\Delta_{nklm}R_{mj}-(i\leftrightarrow j)\right)~. \label{maximal-Q2}
\end{eqnarray}
Here we use the property that $\mathsf{(A-A^{\tau}})/2$ is the antisymmetric part of a matrix $\mathsf{A}$.
In general the resultant current $\mathbf{j}^{(1)}_{on}$ is not divergenceless.

The current $\mathbf{j}^{(1)}_{off}$ flowing off the equiprobability surface is given from the remaining part in $\mathbf{f}^{(1)}_{res}$ after the maximal extraction of $\mathsf{Q}^{(1)}$. We find the velocity field of it as
\begin{eqnarray}
\lefteqn{[\mathbf{f}_{off}]_{i}=\mathsf{[D+Q^{(0)}]}_{ij}h_{j} }\nonumber\\
&&+{1\over2}\left[\left\{\mathsf{(D+Q^{(0)})U^{(1)}R}+\mathsf{RU^{(1)}(D-Q^{(0)})}\right\}
\mathsf{U}^{(0)}\right]_{ik}x_{k} \nonumber\\
&&+\left({1\over 2}G_{ikl}+{1\over 2}G_{jkl'}R_{l'i}U^{(0)}_{jl}
+{1\over 4}[\mathsf{D+Q^{(0)}}]_{im}\Gamma_{mkl}    \right.\nonumber\\
&&\left. +{1\over 4}[\mathsf{D+Q^{(0)}}]_{jm}\Gamma_{mkl'} R_{l'i}U^{(0)}_{jl}\right)
x_{k}x_{l} \nonumber\\
&&+\left({1\over 2}H_{iklm}+{1\over
2}H_{jklm'}R_{m'i}U^{(0)}_{jm}+ {1\over 12}[\mathsf{D+Q^{(0)}}]_{in}\Delta_{nklm}\right.\nonumber\\
&&  \left. + {1\over 12}[\mathsf{D+Q^{(0)}}]_{jn}\Delta_{nklm'}R_{m'i}U^{(0)}_{jm} \right)x_{k}x_{l}x_{m}~. \label{maximal-feff}
\end{eqnarray}
The resultant current $\mathbf{j}^{(1)}_{off}$ may not be divergenceless, while the total current of the first order, $\mathbf{j}^{(1)}=\mathbf{j}^{(1)}_{on}+\mathbf{j}^{(1)}_{off}$, is expected to be divergenceless. As observed in an example in the next section, 
$\mathbf{j} ^{(1)}$ shows interesting behaviors. In particular the current yields circulation at multiple centers which are not coincident with either the probability maximum or the fixed point.   

\section{Motion in a Two-Dimensional Potential Well\label{section6}}

Let us consider a motion in a two-dimensional double well potential $V(x,y)$ for $x_1=x$ and $x_2=y$. The double well potential is given by
\begin{equation}
V(x,y)= {k_1 \over 2a^2}x^2(x-a)^2+ {k_2\over 2}y^2~.\label{double-well}
\end{equation}
There are three fixed points where $\bm{\nabla} V=0$: $(0,0)$ (stable), $(a/2,~0)$ (saddle), $(a,~0)$ (stable). The diffusion matrix is taken to have nonzero off-diagonal element $D_{12}$.

We carry out the perturbation theory around $(0,0)$. Note that the force matrix,  given by $F_{ij}=-\partial^2 V/\partial
x_i\partial x_j$, is symmetric. However, noise correlation leads the system to nonequilibrium. In zeroth order we get
\begin{equation}
\mathsf{F}^{(0)}=-\left(
\begin{array}{cc}
  k_1 & 0 \\
  0 & k_2
\end{array}\right)~.
\end{equation}
The coefficients of nonlinear terms are given by
\begin{equation}
H_{ijk}=f_3 \delta_{i1}\delta_{j1}\delta_{k1}~, 
G_{ijkl} =f_4\delta_{i1}\delta_{j1}\delta_{k1}\delta_{l1}~,
\end{equation}
where 
\[
f_3={3k_1\over a}~,~~f_4= -{2k_1\over a^2}~.
\] 
The full force matrix is given by
\begin{equation}
\mathsf{F}=\left(
\begin{array}{cc}
  -k_1+2f_3 x_1+3f_{4}x_1^2/2 & 0 \\
  0 & -k_2
\end{array}\right)~.
\end{equation}
Then the condition for DB, $\mathsf{FD-DF^{\tau}=0}$, gives
\begin{equation}
(-k_1+k_2+2f_{3}x_1+3f_{4}x_1^2)D_{12}=0~.
\end{equation}
The DB is violated for $D_{12}\neq 0$. 

In proceeding the perturbation theory, we need to carry out multiple integrals for functions given in terms of matrices $\alpha(\omega)$, $\beta(\omega)$ defined in
Eq.~(\ref{alpha-beta}). We find
\begin{eqnarray}
\mathsf{\alpha}(\omega)&=&\left(
\begin{array}{cc}
  {1\over -i\omega+k_1} & 0 \\
  0 & {1\over -i\omega +k_2}
\end{array}\right)~, \\
\mathsf{\beta}(\omega)&=&\left(
\begin{array}{cc}
  {D_{11}\over (-i\omega+k_1)(i\omega+k_1)} & {D_{12}\over (-i\omega+k_1)(i\omega+k_2)} \\
  {D_{12}\over (-i\omega+k_2)(i\omega+k_1)} & {D_{22}\over (-i\omega+k_2)(i\omega+k_2)}
\end{array}\right)~.
\end{eqnarray}
One can find $\mathsf{U}^{(0)}$ and $\mathsf{Q}^{(0)}$ from the integral representations in Eqs.~(\ref{U-inverse-integral}) and (\ref{Q-integral}):
\begin{equation}
\mathsf{R}=\left(
\begin{array}{cc}
  {D_{11}\over k_1} & {2D_{12}\over k_1+k_2} \\
  {2D_{12}\over k_1+k_2} & {D_{22}\over k_2}
\end{array} \right)~,
\end{equation}
and
\begin{equation}
Q^{(0)}_{12}={k_1-k_2 \over k_1+k_2}D_{12}~\label{Q0}
\end{equation}

For simplicity we consider the case where $k_1=k_2=k$, $D_{11}=D_{22}=d$, $D_{12}= \epsilon d$. Then  we find 
\begin{equation}
\mathsf{Q^{(0)}=0~,~~ 
U^{(0)}}=\frac{k}{d(1-\epsilon^2)}
\left(
\begin{array}{cc}
  1 & -\epsilon \\
  -\epsilon & 1
\end{array} \right)~,
\end{equation}

After carrying out multiple residue integrals, we find
\begin{eqnarray}
\Theta_{11}^{11}&=&{6\over {d(1-\epsilon^2)}}~,\\
\Theta_{21}^{11}&=&\Theta_{12}^{11}=-{{3\epsilon}\over {2d(1-\epsilon^2)}}~,
\end{eqnarray}
and
\begin{eqnarray}
\Gamma_{111}&=& -{{2f_{3}}\over {d(1-\epsilon^2)} }~, \label{Gamma111}\\
\Gamma_{211} &=&\Gamma_{121}=\Gamma_{112}={{2\epsilon f_{3}}\over {3d(1-\epsilon^2)} }~, \label{Gamma211}
\end{eqnarray}
and
\begin{eqnarray}
\Delta_{1111}&=& -{{6f_{4}}\over {d(1-\epsilon^2)} }~,\\
\Delta_{2111}&=&\Delta_{1211}=\Delta_{1121}=\Delta_{1112}\nonumber\\
&=&{{3\epsilon f_{4}}\over {2d(1-\epsilon^2)} }~.
\end{eqnarray}

Now $\mathsf{U^{(1)}}$ can be found as
\begin{eqnarray}
U^{(1)}_{11}&=&-{{3\epsilon^2f_{4}}\over {2k(1-\epsilon^2)}}~,\\
U^{(1)}_{21}&=&{{3\epsilon f_{4}}\over {4k(1-\epsilon^2)}}~,\\
U^{(1)}_{22}&=&0~.
\end{eqnarray}
The coefficients $h_{i}$ for the linear term in $\Phi$ can also be found as
\begin{eqnarray}
h_{1}&=&-{2\epsilon^2 f_{3}\over 3(1-\epsilon^2)}~,\\
h_{2}&=& {2\epsilon f_{3} \over 3k(1-\epsilon^2)}~.
\end{eqnarray}

The shift of probability maximum from the fixed point is found as
\begin{eqnarray}
x_{max}&=&-R_{1i}h_{i}=0~,\label{shift1}\\
y_{max}&=&-R_{2i}h_{i}=-{2\epsilon d f_{3}\over 3 k^2}~.\label{shift2}
\end{eqnarray}
The shift is made perpendicular to $x$ direction where the nonlinear force is applied. It is somewhat contrary to our expectation. There may be an effective transverse force 
due to noise correlation. Note that the shift depends on $D_{12}$ and $f_3$. It is a novel phenomenon due the combination of noise correlation and 
nonlinearity in the drift force.  

The average position $\langle \mathbf{x}\rangle$ can be found from Eq.~(\ref{first-mom}),
\begin{equation}
\langle x_i \rangle = -F^{(0)}_{i1}f_{3}[U^{(0)}]^{-1}_{11}={{df_{3}}\over{ k^2}}\delta_{i1}~.
\end{equation}
Note that it is in $x$ direction, perpendicular to $\mathbf{x}_{max}$.

The potential landscape function is then given as
\begin{eqnarray}
\Phi(x,y) &=& \frac{k}{2d\kappa}(x^2+y^2-2\epsilon xy)+\frac{2\epsilon f_3}{3\kappa}(-\epsilon x+y)\nonumber\\
&&+\frac{3\epsilon f_4}{4k\kappa}(-\epsilon x^2+xy)+\frac{f_3}{3d\kappa}(-x^3+\epsilon x^2y)\nonumber\\
&& +\frac{f_4}{4d\kappa}(-x^4+\epsilon x^3 y)~,
\label{landscape}
\end{eqnarray}
where $\kappa=1-\epsilon^2$. Note that the perturbation theory is valid for small $d$ where we can treat $\mathbf{x}$ to be ${\cal O}(\sqrt{d})$. Therefore the first order perturbation theory 
with respect to nonlinear coefficient $f_3$, $f_4$ is equivalent to the expansion of $\Phi$ up to the first order in $d$. This implies that $f_3$, $f_4$ are not necessarily small if $d$ is taken to be small. 

The antisymmetric matrix in the first order is found as 
\begin{equation}
Q_{12}^{(1)}=-\frac{\epsilon d}{2k}\left(f_{3}x+f_{4}x^2\right)~.
\end{equation}
As noticed it depends on $\mathbf{x}$. The resultant current $\mathbf{j}^{(1)}_{on}=-(\mathsf{Q}^{(1)}\cdot\bm{\nabla}\Phi^{(0)})\rho$ flows on the equiprobability surface, but is not divergenceless. 

The velocity field for the current $\mathbf{j}^{(1)}_{off}$ flowing off the equiprobability surface is found as 
\begin{eqnarray}
[\mathbf{f}^{(1)}_{off}]_x&=&{3\epsilon d f_{4}\over 4k\kappa}\left(-\epsilon x+y)\right)
-{\epsilon f_{3}\over 6\kappa} \left( \epsilon x^2-xy\right)
\nonumber\\
&&-{\epsilon f_{4}\over 4\kappa}\left( \epsilon x^3-x^2y\right)~,\\ \label{f_1_max} 
[\mathbf{f}^{(1)}_{off}]_y &=&{2\epsilon df_{3}\over 3k}
+{3\epsilon d f_{4}\over 4k\kappa}\left((1-2\epsilon^2)x+\epsilon y \right) \nonumber\\
&&-{\epsilon f_{3}\over 6\kappa}\left(x^2-\epsilon xy\right) 
-{\epsilon f_{4}\over 4\kappa}\left(x^3-\epsilon x^2y\right)~.
\label{f_2_max}
\end{eqnarray}

The velocity field for the total current $\mathbf{j}^{(1)}$ in the first order can be found as
\begin{eqnarray}
[\mathbf{f}^{(1)}_{res}]_x&=&{3\epsilon d f_{4}\over 4k\kappa}\left(-\epsilon x+y)\right)
-{2\epsilon f_{3}\over 3\kappa} \left( \epsilon x^2-xy\right)\nonumber\\
&&-{3\epsilon f_{4}\over 4\kappa}\left( \epsilon x^3-x^2y\right)~,\\ \label{f_1} 
[\mathbf{f}^{(1)}_{res}]_y &=&{2\epsilon df_{3}\over 3k}
+{3\epsilon d f_{4}\over 4k\kappa}\left((1-2\epsilon^2)x+\epsilon y \right) \nonumber\\
&&-{2\epsilon f_{3}\over 3\kappa}\left(x^2-\epsilon xy\right)
-{3\epsilon f_{4}\over 4\kappa}\left(x^3-\epsilon x^2y\right)~.\label{f_2}
\end{eqnarray}
One can show $\mathbf{j}^{(1)}$ is divergenceless, i.e., $\bm{\nabla}\cdot \mathbf{j}^{(1)}=0$, as expected from the steady state condition. Therefore we expect that it yields circulation off the equiprobability surface. The current vanishes at the center of circulation. Surprisingly there are two centers at $(x_c, y_c)$:
\begin{equation}
x_c=\pm \sqrt\frac{d}{k}~,~~y_c=\mp\epsilon\sqrt\frac{d}{k}~,\label{center}
\end{equation}  
where we assume $d$ is small, which is the criterion for the perturbation theory.

\section{Numerical Studies\label{section7}}

In order to confirm the results obtained from the perturbation theory we solve the stochastic differential equation (\ref{stochastic}) numerically. We consider the example in the last section. Let $\{t_n; n=0,\ldots, N\}$ be discrete time steps with interval $\Delta t$. We write $x(t_n)=x_n$, $y(t_{n})=y_n$. Then $x_n$ and $y_n$ are updated as     
\begin{eqnarray}
x_{n}&=&x_{n-1}+f_{x,n-1}\Delta t +\left( \sqrt{1-\epsilon^2}\xi_1+\epsilon\xi_2\right)\sqrt{2d\Delta t}~,\nonumber\\
y_{n}&=& y_{n-1}+f_{y,n-1}\Delta t +\xi_2\sqrt{2d\Delta t}~,
\end{eqnarray}
where $f_{x,n-1}=-kx_{n-1}+f_3 x^2_{n-1} +f_4 x^3_{n-1}$, $f_{y,n-1}=-k y_{n-1}$. $\xi_{1,2}$ are random numbers chosen independently from $\{-1, 1\}$ at each time step. We execute the simulation up to $N=10^9$ steps. 

\begin{figure}[t]
\includegraphics*[width=\columnwidth]{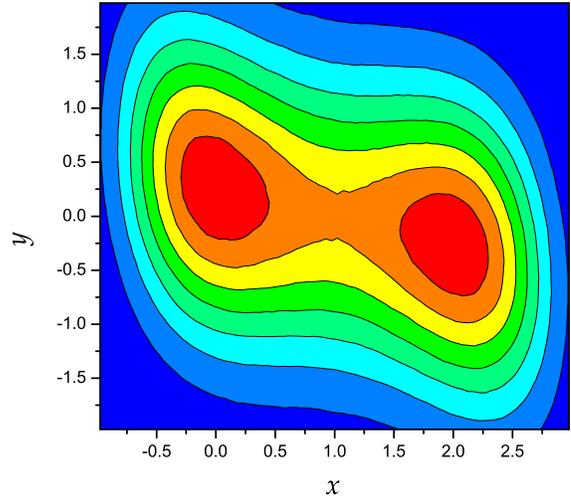}
\caption{\label{fig1} The contour plot of the PDF from the numerical simulation for $d=1.0$, $\epsilon=-0.3$, $a=2.0$, $k=1.0$. }
\end{figure}

\begin{figure}[t]
\includegraphics*[width=\columnwidth]{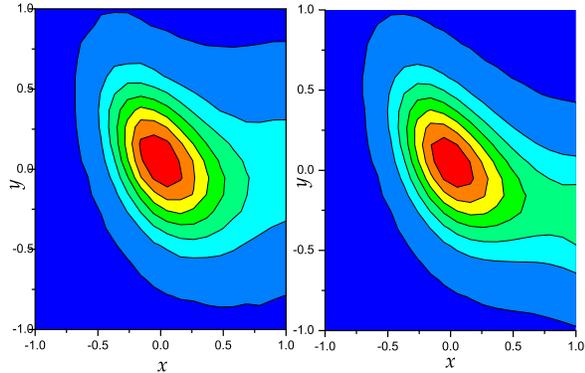} 
\caption{Contour plots for PDF from theory and simulation for $d=1.0$, $\epsilon=-0.5$, $a=2.0$, $k=1.0$.}
\label{fig2}
\end{figure}

The shift of the PDF maximum is proportional to $d$, as shown in Eq.~(\ref{shift2}).  In order to show the shift clearly we take $d=1.0$, that is beyond the perturbation theory. Fig.~\ref{fig1} shows the contour plot for the PDF obtained from the simulation.  We use a negative value of $D_{12}$, so the shift from $(0,0)$ is in $+y$ direction, agreeing with Eq.~(\ref{shift2}). Around the saddle point $(1,0)$, where $f_{111}=0$, there is no shift, which can be seen from Eq.~(\ref{shift}). The shift from the other fixed point $(2,0)$ is in opposite direction, as can be expected from the symmetry of the potential well.
Fig.~\ref{fig2} shows the two contour plots of the PDF obtained from the perturbation theory and the numerical simulation. We take $d=0.1$ for which the perturbation theory is relevant. The two plots seem to be well coincident to each other near $(0,0)$, maybe within the window of side $\sim\sqrt{d}$, that is about $0.3$.  
\begin{figure}[t]
\includegraphics*[width=\columnwidth]{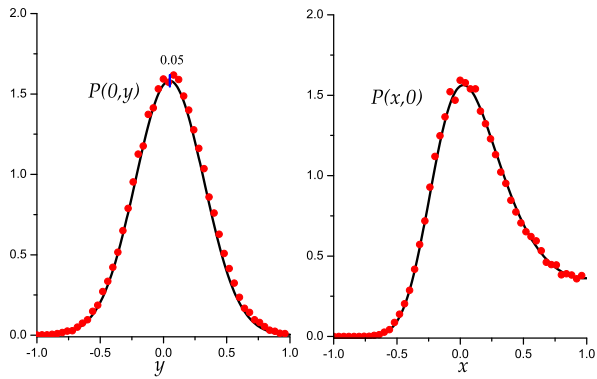}
\caption{PDF along $x=0$ and $y=0$ lines through the origin for the same values of parameters as in Fig.~\ref{fig2}. The maximum of $P(0,y)$ is estimated to be $0.05$, which is the value of the shift of the PDF maximum from the origin. The solid line is for the theory and the red circles for the simulation}
\label{fig3}
\end{figure}
In Fig.~\ref{fig3} we present the PDF along the two lines, $x=0$ and $y=0$, through the origin. The maximum of $P(0,y)$ is the shifted value of the PDF maximum. It is estimated from the perturbation theory as $0.05$. The solid line is the plot from the theory and the scattered circles from the simulation. The figure shows a good agreement between the two methods.  

\begin{figure}
\includegraphics*[width=\columnwidth]{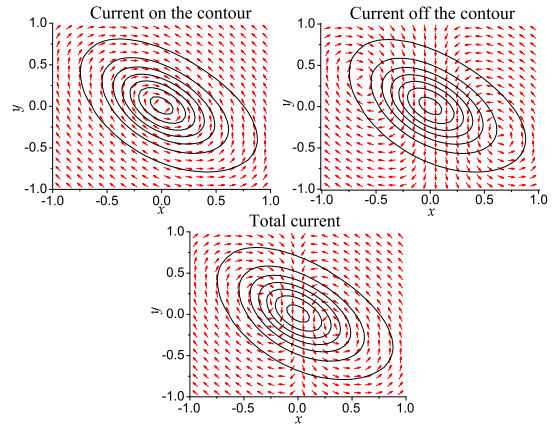}
\caption{NESS Current. The current flowing on the contour (top left), the current off the contour (top right), the total current (bottom). We take the same values of parameters in the last figure except for $a=10.0$. There appear two centers of circulation for the total current at $(\pm 0.316, \mp 0.158)$ }
\label{fig4}
\end{figure}
Fig.~\ref{fig4} shows the NESS current drawn on the contour lines. For this example the current in the zeroth order vanishes. We take a larger value for $a$. The larger $a$, the smaller $f_3$ and $f_4$. Then the perturbation theory will give a more accurate estimation in the region far from the origin. The figure on the top left shows  $\mathbf{j}^{(1)}_{on}=-(\mathsf{Q}^{(1)}\cdot\bm{\nabla}\Phi^{(0)})\rho$ flowing on the contour line (equiprobability surface),  the figure on the top right shows $\mathbf{j}^{(1)}_{off}=\mathbf{f}_{off}\rho$ flowing off the contour line, and the figure at the bottom shows the total current $\mathbf{j}^{(1)}=\mathbf{j}^{(1)}_{on}+\mathbf{j}^{(1)}_{off}$. The current on the contour line shows a cut (discontinuity line), at $x=0$; the current flows into the cut for $y>0$ and flows out from the cut for $y<0$. It manifests that the current is not divergenceless, as expected from the dependence of $\mathsf{Q}^{(1)}$ on $\mathbf{x}$. The current $\mathbf{j}^{(1)}_{off}$ is not divergenceless, not clearly seen in the figure. It can be observed that there is a local flux flowing down through the origin, filling up the discontinuity of $\mathbf{j}^{(1)}_{on}$. As a result, the total current becomes divergenceless while circulation splits into two parts. It is not an artifact of the perturbation theory, for the PDF is accurate in the region near the centers of circulation. For $d=0.1$ and $k=1.0$, they are located at $(\pm 0.316, \mp 0.158)$, estimated from Eq.~(\ref{center}).  The circulation takes place across the contour line, which is an important characteristics for the NESS together with the shift of the PDF maximum from the fixed point. However, it is quite surprising that there appear several centers of circulation. 

\section{Discussion\label{section8}}

We have investigated the potential landscape and the NESS current for the localized stochastic system driven by the nonlinear drift force and in contact with nonuniform noises. Unlike the extended system the induced NESS current is circulating. For the linear drift force it was found to circulate around the equiprobability surface~\cite{kat} with the probability maximum at the fixed point of the drift force. In this study we have found that the NESS current flows off the equiprobability surface and circulates around multiple centers. The PDF maximum is found to shift from the fixed point of the drift force. For an example we take a familiar potential well yielding a conservative force. Difference is made only by noise correlation. However, it turns out that the resultant potential landscape, having multiple centers of current circulation, is rather drastically deformed from the equilibrium shape. Circulation is found to split into two parts and the boundary current fills up the discontinuity of the current flowing on the equiprobability surface. However, further investigation is needed for a more plausible explanation based on the physical origin. Noise induced transverse force might be a possible origin, which is not confirmed yet. They are novel phenomena caused by the combination of two causes, the nonlinearity of the drift force and the nonuniformity of high dimensional noises. Interestingly they are basically equivalent to the two ingredients for the noise induced transportation current found for the extended periodic system: the ratchet type potential (nonlinear force) and the additional noise (high dimensional nonuniform noises)~\cite{prost,doering}. Our finding is based on the first order perturbation theory that is valid in the low noise limit. It is confirmed by numerical simulations.

It is interesting to realize the NESS with a circulating current experimentally, probably in an optical trap experiment in two dimensions. It is a challenging task to produce noises experimentally which are not identically and independently distributed. Force in experiments may be conservative and then the force matrix $\mathsf{F}$ is symmetric. In this case one can still produce a nonequilibrium situation if the principal axes of the diffusion matrix are chosen to be  different from those of $\mathsf{F}$, as suggested by Filliger et al.~\cite{filliger}. 

The fluctuation of nonequilibrium work production has recently been studied for the linear case~\cite{turitsyn,knp1}. There are interesting properties found such as the exponential tail with a power law prefactor of the work distribution function and the dynamic phase transition in the exponent of the prefactor. The properties of the NESS for the nonlinear case are found to be quite different from that for the linear case. In this sense it will be interesting to study the nonlinear case from the point of view of the fluctuation theorem. 

\begin{acknowledgments}
We thank David Thouless, Hyunggyu Park, and Jae Dong Noh for many helpful discussions. We also thank Jacques Prost for stimulating discussion about the relation of our work with the previous work by his group on the transportation current in asymmetric pumping~\cite{prost}. This work was supported by Mid-career Researcher Program through NRF grant (No.~2010-0026627) funded by the MEST. P. A. appreciates the partial support by China National 973 Projects No. 2007CB914700 and No. 2010CB529200.
\end{acknowledgments}


\begin{thebibliography}{99}
\bibitem{evans} D. J. Evans, E. G. D. Cohen, and G. P. Morriss, Phys. Rev. Lett {\bf 71}, 2401 (1993).
\bibitem{evans-searles1} D. J. Evans and D. J. Searles, Phys. Rev. E {\bf 50}, 1645 (1994); Phys. Rev. E {\bf 52}, 58093 (1995); Phys. Rev. E {\bf 53}, 5808 (1996).
\bibitem{gallavotti1} G. Gallavotti and E. G. D. Cohen, Phys. Rev. Lett. {\bf 74}, 2649 (1995); J. Stat. Phys. {\bf 80}, 931 (1995).
\bibitem{gallavotti2} G. Gallavotti, Phys. Rev. Lett. {\bf 77}, 4334 (1996).
\bibitem{crooks1} G. E. Crooks, J. Stat. Phys. {\bf 90}, 1481 (1998).
\bibitem{kurchan1} J. Kurchan, J. Phys. A: Math. Gen. {\bf 31},3719 (1998).
\bibitem{lebowitz1} J. L. Lebowitz and H. Spohn, J. Stat. Phys. {\bf 95}, 333 (1999)
\bibitem{maes} C. Maes, J. Stat. Phys. {\bf 95}, 367 (1998)
\bibitem{hatano-sasa} T. Hatano and S. Sasa, Phys. Rev. Lett. 86 3463 (2001).
\bibitem{cugliandolo} L. Cugliandolo, J. Kurchan, L. Peliti, Phys. Rev. E {\bf 55} 3898 (1997).
\bibitem{eyink} G. L. Eyink, J. L. Lebowitz, and H. Spohn, J. Stat. Phys. {\bf 83}, 385 (1998).
\bibitem{bellon} L. Bellon, S. Ciliberto, and C. Laroche, Europhys. Lett {\bf 53} 511 (2001).
\bibitem{harada-sasa} T. Harada and S. Sasa, Phys. Rev. Lett. {\bf 95}, 130602 (2005).
\bibitem{chetrite1} R. Chetrite and S. Gupta, J. Stat. Phys. {\bf 143}, 543 (2011).
\bibitem{prost} J. Prost, J.-F. Chauwin, L. Peliti, and A. Ajdari, Phys. Rev. Lett. {\bf 72}, 2652 (1994).
\bibitem{doering} C. Doering, W. Horsthemke, and J. Riordan, Phys. Rev. Lett. {\bf 72}, 2984 (1994).
\bibitem{derrida1} B. Derrida, J. L . Lebowitz, and E. R. Speer, Phys. Rev. Lett. {\bf 87}, 150601 (2001).
\bibitem{bodineau1} T. Bodineau and B. Derrida, Phys. Rev. Lett. {\bf 92}, 180601 (2004).
\bibitem{kat} C. Kwon, P. Ao, and D. Thouless, Proceed. Nat. Acad. Sci. {\bf 102}, 13029 (2005).
\bibitem{bodineau2} T. Bodineau and B. Derrida, C. R. Physique {\bf 8}, 540 (2007).
\bibitem{derrida2} B. Derrida, J. Stat. Mech., P07023 (2007).
\bibitem{knp1} C. Kwon, J. D. Noh, and H. Park, Phys. Rev. E {\bf 83}, 061145 (2011).
\bibitem{garnier} N. Garnier and S. Ciliberto,. Phys. Rev. E {\bf 71}, 60101 (2005).
\bibitem{douarche1} F. Douarche, S. Joubaud, N. B.  Garnier, A. Petrosyan, and S. Ciliberto, Phys. Rev. Lett. {\bf 97}, 140603 (2006).
\bibitem{visco} P. Visco, J. Stat. Mech., P06006 (2006).
\bibitem{komatsu} T. S. Komatsu and N. Nakagawa, Phys. Rev. Lett. {\bf 100}, 030601 (2008).
\bibitem{turitsyn} K. Turitsyn, M. Chertkov, V. Y. Chernyak, and A. Puliafito, Phys. Rev. Lett. {\bf 98}, 180603 (2007).
\bibitem{chetrite2} R. Chetrite and K. Gawedzki, Commun. Math. Phys. {\bf 282}, 469 (2008).
\bibitem{jarzynski1} C. Jarzynski, Phys. Rev. Lett. {\bf 78}, 2690 (1997); Phys. Rev. E {\bf 56}, 5018 (1997); J. Stat. Phys. {\bf 98}, 77 (2000).
\bibitem{crooks2} G. E. Crooks, Phys. Rev. E {\bf 61}, 2361 (2000).
\bibitem{mazonka} O. Mazonka and C. Jarzynski, arXiv: 9912121 (1999).
\bibitem{zon} R. van Zon and E. G. D. Cohen, Phys. Rev. Lett. {\bf 91}, 110601 (2003).
\bibitem{taniguchi} T. Taniguchi and E. G. D. Cohen,  J. Stat. Phys. {\bf 126} 1 (2007).
\bibitem{angel} A. Angel, Phys. RTev. E. {\bf 80}, 021120 (2009).
\bibitem{knp2} C. Kwon, J. D. Noh, and H. Park, {\it Nonequilibrium fluctuation due to the time varying harmonic potential beyond the overdamped limit}, to be submitted (2011).
\bibitem{zhu} X.-M. Zhu, L. Yin, L. Hood, and P. Ao,  Funct. Integr. Genom. {\bf  4}, 185 (2004).
\bibitem{ao1} P. Ao, J. Phys. A {\bf 37}, L25 (2004). 
\bibitem{yin} L. Yin and P. Ao, J. Phys. A {\bf 39}, 8593 (2006).
\bibitem{ao2} P. Ao, C. Kwon, and H. Qian, Complexity {\bf 12}, 19 (2007).
\bibitem{ao3} P. Ao, Commun. Theor. Phys. {\bf 49}, 1073 (2008).
\bibitem{tailleur} J. Tailleur, J. Kurchan, and V. Lecomte, Phys. Rev. Lett. {\bf 99}, 150602 (2007).
\bibitem{graham} R. Graham and T. T\'{e}l, Phys. Rev. Lett. {\bf 52}, 9 (1984); Phys. Rev. A {\bf 33}, 1322 (1986).
\bibitem{wang} J. Wang, L. Xu, and E. Wang, Proceed. Nat. Acad. Sci. {\bf 105}, 12271 (2008).
\bibitem{bertini} L. Bertini and G. Di Gesu, arXiv:1004.237 (2010).
\bibitem{risken} H. Risken, {\it The Fokker-Planck Equation: methods of solution and applications}, 2nd edition (Springer-Verlag, Berlin), pp.145-153 (1989).
\bibitem{puglisi} A. Puglisi and D. Villamaina, Europhys. Lett. {\bf 88}, 30004 (2009).
\bibitem{filliger} R. Filliger, P. Reimann, Phys. Rev. Lett. {\bf 99}, 230602 (2007).
\bibitem{qian} H. Qian, Phys. Rev. Lett. {\bf 81}, 3063 (1998).
\end{thebibliography}
\end{document}